\documentclass[english,pra, twocolumn]{revtex4-2}
\usepackage[T1]{fontenc}
\usepackage[latin9]{inputenc}
\usepackage{amsmath}
\usepackage{amssymb}
\usepackage{graphicx}

\makeatletter
\usepackage[colorlinks = true,
            linkcolor =red,
            urlcolor  =magenta,
            citecolor = blue,
            anchorcolor = blue]{hyperref}
\begingroup

\endgroup

\makeatother

\usepackage{babel}
\begin{document}
\title{Kirkwood-Dirac Quasiprobability as a Universal Framework for Quantum
Measurements Across All Regimes}
\author{Bo Zhang}
\author{Yusuf Turek}
\email{yusufu1984@hotmail.com}

\affiliation{School of Physics,Liaoning University,Shenyang,Liaoning 110036,China}
\date{\today}
\begin{abstract}
The question of when the Kirkwood-Dirac (KD) quasiprobability serves
as the most appropriate description for quantum measurements has remained
unresolved, particularly across different measurement strengths. While
known to generate anomalous weak values in the weak measurement regime
and to reduce to classical probabilities under projective measurement,
the physical mechanism governing its continuous transformation has
been lacking. Here we demonstrate that the KD quasiprobability provides
a general framework for all measurement regimes by identifying pointer-induced
decoherence as the universal mechanism controlling this transition.
We show that the decoherence factor $F(t)$ simultaneously quantifies
the loss of quantum coherence and interpolates the measurement strength
from weak to strong. Within this framework, the KD quasiprobability
naturally deforms from its full complex form---governing weak values---to
the real, non-negative Wigner formula describing projective measurements,
while maintaining informational completeness throughout the transition.
Our work resolves the fundamental question of the KD distribution's
applicability by establishing it as the universal framework that seamlessly
connects all quantum measurement regimes through a physically transparent
decoherence pathway.
\end{abstract}
\maketitle

\section{\label{sec:1}Introduction}

Quantum measurement theory has long been divided into two apparently
distinct paradigms. On one hand, strong (projective) measurements
collapse the system\textquoteright s state and yield definite outcomes
consistent with the Born rule \citep{Lueders1951,Neumann1932}. On
the other, weak measurements minimally disturb the system and can
give rise to complex-valued \textquotedblleft weak values\textquotedblright{}
that may lie far outside an observable\textquoteright s eigenvalue
spectrum \citep{Aharonov1988,Dressel2014}. A unified theoretical
description that bridges these extremes---and, importantly, the continuum
of measurement strengths in between---has remained an outstanding
challenge.

Quasiprobability distributions offer a natural language for describing
quantum states beyond classical phase-space representations. Among
them, the Kirkwood--Dirac (KD) distribution \citep{Johansen2004,Dirac1945,Arvidsson-Shukur2024},
defined with respect to two noncommuting observables, is distinguished
by its informational completeness and direct connection to weak values
and quantum contextuality \citep{Hofmann2012,Wagner2023,Pusey2014}.
It is known to produce anomalous weak values in the weak-measurement
regime and to reduce to the non-negative Wigner formula under projective
measurement. However, a fundamental gap remains: what is the physical
mechanism that governs the continuous transformation of the KD distribution
between these limits, and when is it the most appropriate description
for general quantum measurements?

In this work, we demonstrate that the KD quasiprobability itself constitutes
this structure. We show that the physics of the measurement transition---including
the universal role of the decoherence factor $F(t)$ ---emerges directly
from a smooth deformation of the KD distribution. This deformation
continuously interpolates between the full complex-valued form governing
weak values and the real, non-negative Wigner formula appropriate
for projective measurements, all while maintaining informational completeness.

While recent work has identified the overlap of pointer states---quantified
by a single time-dependent factor $F(t)$---as a universal mechanism
bridging weak and strong measurements \citep{Song2024,Schlosshauer2007}.
a fundamental question remains: What is the natural mathematical structure
that encapsulates this transition while preserving informational completeness
across all measurement regimes?

Moreover, we establish that the non-classicality of the KD distribution,
quantified by a suitable measure, decays linearly with $F(t)$, thereby
linking the formal transition parameter directly to the loss of quantum
coherence \citep{Streltsov2017}. This elevates the KD distribution
from a passive quasiprobability representation to the central mathematical
framework from which the phenomenology of weak, intermediate, and
strong measurements naturally emerges---a step that provides both
a deeper unification and a quantitative tool for analyzing quantumness
across the entire measurement landscape.

We organized the paper as follows: In Sec. \ref{sec:2}, we introduce
the Kirkwood-Dirac distribution and establish its fundamental properties
as an informationally complete representation of quantum states. We
then demonstrate how it unifies the description of normal expectation
values, conditional expectation values (ABL rule), and weak values.
In Sec. \ref{sec:3}, we develop a dynamical model of measurement-induced
decoherence and show how the decoherence factor $F(t)$ continuously
interpolates between measurement regimes through the time-dependent
KD distribution. In Sec. \ref{sec:4}, we analyze how the non-classicality
of the KD distribution decays linearly with $F(t)$, providing a quantitative
measure of quantumness across the measurement transition. Finally,
we conclude in Sec. \ref{sec:5}. Throughout this paper, we set $\hbar=1$.

\section{\label{sec:2}Definition and basis properties }

In this section, based on the Kirkwood-Dirac distribution, a general
framework for quantum measurement is constructed. First, the KD distribution
is defined, showing that it can represent quantum states and provide
marginal probabilities. Then, the real part, imaginary part, and correction
terms of the KD distribution are decomposed to explain the source
of non-classicality. Finally, based on the KD distribution, normal
expectation values, conditional expectation values, and weak values
are uniformly expressed, and the applications and physical meaning
of the KD distribution in scenarios such as mixed states and non-projective
measurements are analyzed, achieving integration of quantum measurement
scenarios.

\subsection{\label{subsec:2-1}Kirk wood-Dirac distribution }

The Kirkwood-Dirac (KD) quasiprobability distribution provides a fundamental
and informationally complete representation of quantum states with
respect to pairs of non-commuting observables \citep{Dirac1945,Langrenez2023,Kirkwood1933,Arvidsson-Shukur2024}.
Here, we establish its formal definition and core mathematical properties,
synthesizing and extending the operational insights from Johansen's
projective measurement framework \citep{Johansen2008,Johansen2007}.

Consider a quantum system associated with a finite-dimensional Hilbert
space $\mathcal{H}$ of dimension $d$. Let the $A$ and $F$ be two
non-degenerate observales, with their complete sets of orthogonal
eigenstates denoted by $\{\vert a_{i}\rangle\}_{i=1}^{d}$ and $\{\vert f_{j}\rangle\}_{j=1}^{d}$,
respectively. Their spectral decompositions are: 
\begin{align}
A & =\sum_{i=1}^{d}a_{i}\Pi_{a_{i}},\ \ F=\sum_{j=1}^{d}f_{j}\Pi_{f_{j}},
\end{align}
where $\Pi_{a_{i}}=\vert a_{i}\rangle\langle a_{i}\vert$ and $\Pi_{f_{j}}=\vert f_{j}\rangle\langle f_{j}\vert$
are the rank-one projection operators onto the respective eigenstates. 

For a given quantum state $\rho$, the KD distribution with respect
to the eigenbases of $A$ and $F$ is defined as 
\begin{align}
P_{ij}(\rho) & =Tr\left(\rho\Pi_{a_{i}}\Pi_{f_{j}}\right)=\langle f_{j}\vert\rho\vert a_{i}\rangle\langle a_{i}\vert f_{j}\rangle.
\end{align}
An alternative, mathematically equivalent form is often used: 
\begin{equation}
Q_{ij}(\rho)=Tr\left(\Pi_{a_{i}}\rho\Pi_{f_{j}}\right)=\langle a_{i}\vert\rho\vert f_{j}\rangle\langle f_{j}\vert a_{i}\rangle,
\end{equation}
noting that $Q_{ij}(\rho)=P_{ij}^{\ast}(\rho)$. In this work,we adopt
$Q_{ij}(\rho)$ as our primary definition of analysis. 

A pivotal property of the KD distribution is its informational completeness.
Provided the two bases are mutually non-orthogonal ($\langle f_{j}\vert a_{i}\rangle\neq0$,
for all $i,j$), the density operator $\rho$ can be uniquely and
fully reconstructed from the KD distribution \citep{Johansen2007}:

\begin{align}
\rho & =\sum_{i,j}Q_{ij}(\rho))\frac{\vert a_{i}\rangle\langle f_{j}\vert}{\langle f_{j}\vert a_{i}\rangle}.
\end{align}
 This reconstruction formula demonstrates that the set $\{\frac{\vert a_{i}\rangle\langle f_{j}\vert}{\langle f_{j}\vert a_{i}\rangle}\}$
forms a non-orthogonal operator basis for the space of linear operators
on $\mathcal{H}$, dual to the projector products $\Pi_{a_{i}}\Pi_{f_{j}}$.
The condition $\langle f_{j}\vert a_{i}\rangle$ is equivalent to
the observables $A$ and $F$ being complementary, i.e., sharing no
common eigenstates. In finite dimensions, a stronger condition often
employed is that the bases are mutually unbiased bases (MUBs) \citep{Scott2006,Rastegin2025},
where $\langle f_{j}\vert a_{i}\rangle=1/\sqrt{d}$. 

Despite being complex-valued, the KD distribution satisfies key properties
reminiscent of a classical joint probability distribution. Its marginal
sums yield the correct Born-rule probabilities:\begin{subequations}

\begin{align}
\sum_{j}Q_{ij}(\rho) & =Tr\left(\rho\Pi_{a_{i}}\right)=p(a_{i}\vert\rho),\label{eq:5}\\
\sum_{i}Q_{ij}(\rho) & =Tr\left(\rho\Pi_{f_{j}}\right)=p(f_{j}\vert\rho),\\
\sum_{ij}Q_{ij}(\rho) & =Tr\left(\rho\right)=1.
\end{align}
 \end{subequations}Here, $p(a_{i}\vert\rho)$ is probability of obtaining
eigvenvalue $a_{i}$ of system observable $A$ when the quantum state
is $\rho$. Similarly, $p(f_{j}\vert\rho)$ is the probability of
getting postselected state $\vert f_{j}\rangle$ when the quantum
state is $\rho$. This marginal consistency is a hallmark of the KD
representation, directly linking it to measurable statistics. However,
the distribution itself is a quasiprobability: its elements $Q_{ij}(\rho)$
can be negative or non-real. This non-positivity is a direct signature
of quantum non-classicality, arising from the incompatibility (non-commutation)
of the observables $A$ and $F$ and the coherence present in the
state $\rho$ \citep{DeBievre2021}. 

Following Johansen's seminal result \citep{Johansen2007}, the KD
distribution can be decomposed in an operationally significant way
that separates a classical-like joint probability from a quantum correction
term which written as: 
\begin{align}
Q_{ij}(\rho) & =Tr\left(\rho\Pi_{a_{i}}\Pi_{f_{j}}\Pi_{a_{i}}\right)+\frac{1}{2}Tr\left(\rho-\rho^{\prime}\right)\Pi_{f_{j}}\nonumber \\
 & -\frac{i}{2}Tr\left(\rho-\rho^{\prime}\right)\Pi_{f_{j}}^{\pi/2}.\label{eq:10}
\end{align}
 Here, $\rho^{\prime}$ is the post-measurement state after a non-selective
projection measurement of characterized by $\{\Pi_{a_{i}},1-\Pi_{a_{i}}\}$,
and its expression given by 
\begin{align}
\rho^{\prime} & =\sum_{i}\Pi_{a_{i}}\rho\Pi_{a_{i}}\nonumber \\
 & =\Pi_{a_{i}}\rho\Pi_{a_{i}}+\left(1-\Pi_{a_{i}}\right)\rho\left(1-\Pi_{a_{i}}\right).
\end{align}
 The first and second terms denote the real part of the KD distribution,
i.e, 
\begin{align}
Re\left[Q_{ij}(\rho)\right] & =Tr\left(\rho\Pi_{a_{i}}\Pi_{f_{j}}\Pi_{a_{i}}\right)+\frac{1}{2}Tr\left(\rho-\rho^{\prime}\right)\Pi_{f_{j}}\nonumber \\
 & =Q_{winger}+Re\left[\sum_{k\neq i}\rho_{ki}\Psi_{ki}\right].
\end{align}
 Here, $\rho_{ki}=\langle a_{k}|\rho|a_{i}\rangle$ and $\Psi_{ki}=\langle a_{i}\vert\Pi_{f_{j}}|a_{k}\rangle$
denote the matrix elements of pre- and post-selected states $\vert\psi\rangle$
and $\vert f_{j}\rangle$ under the reference basis $\{\vert a_{i}\rangle\}$.
The first term, $Q_{wigner}$, is the well-known Wigner formula \citep{Wigner1963}.
It represents the classical-like joint probability for obtaining outcome
$a_{i}$ followed by $f_{j}$ in a sequence of two strong (projective)
measurements.

\begin{align}
Q_{winger} & =Tr\left(\rho\Pi_{a_{i}}\Pi_{f_{j}}\Pi_{a_{i}}\right)\label{eq:12}\\
 & =p(a_{i}\vert\rho)p(f_{j}\vert a_{i})
\end{align}
where $p\left(a_{i}\vert\rho\right)=Tr\left(\rho\Pi_{a_{i}}\right)$and�$p\left(f_{j}\vert a_{i}\right)=Tr\left(\rho_{s}'\Pi_{f_{j}}\right)$.
It is the joint probability $Q_{Winger}$ that for first obtaining
the measurement outcome $a_{i}$ and then postselection $f_{j}$ is
performed on the system for the given initial state $\rho$. Here,
$\rho_{s}'$ is the post-measurement state after a selective projection
measurement, and its expression is$\rho_{s}'=Tr\left(\frac{\Pi_{a_{i}}\rho\Pi_{a_{i}}}{Tr\left(\rho\Pi_{a_{i}}\right)}\right)$

Its marginals are: $\sum_{j}Q_{wigner}=p(a_{i}\vert\rho)$ and $\sum_{i}Q_{wigner}=p(f_{j}\vert\rho^{\prime})$.
The marginal probability of $\Pi_{a_{i}}$ depends on the state of
$\rho$, whereas the marginal probability of $\Pi_{f_{j}}$ is given
by state $\rho'$ after a non-selective measurement. The Wigner formula
describes the properties of the pre- and post-measurement states and
does not contain complete information about the pre-measurement state
$\rho$. The second term is a real quantum modification term. It quantifies
the disturbance caused by the first measurement ($\Pi_{a_{i}}$) on
the statistics of the second observable ($\Pi_{f_{j}}$). This term
is the source of potential negativity in the real part of the KD distribution
(often called the Margenau-Hill distribution \citep{Terletsky1937,Margenau1961})
and vanishes if $\left[\rho,\Pi_{a_{i}}\right]=0$ or if $\left[\Pi_{a_{i}},\Pi_{f_{j}}\right]=0$
\citep{Arvidsson-Shukur2021}. The emergence of anomalous values requires
quantum coherence, and the coherence also can be seen .

The third term of Eq. (\ref{eq:10}) represents the imaginary part
of the KD distribution which given by 

\begin{align}
Im\left[Q_{ij}(\rho)\right] & =-\frac{1}{2}Tr\left(\rho-\rho^{\prime}\right)\Pi_{f_{j}}^{\pi/2}\nonumber \\
 & =Im\left[\sum_{k\neq i}\rho_{ki}\Psi_{ki}\right],\label{eq:11}
\end{align}
where the selectively phase-rotated projector $\Pi_{f_{j}}^{\pi/2}$of
the projector $\Pi_{f_{j}}$ is defined by 
\begin{equation}
\Pi_{f_{j}}^{\pi/2}=R_{i}^{\pi/2}\Pi_{f_{j}}\left(R_{i}^{\pi/2}\right)^{\dagger}
\end{equation}
with $R_{i}^{\pi/2}=\mathbb{I}+(e^{i\pi/2}-1)\Pi_{a_{i}}$. This modification
term also also quantifies disturbance, accessed via a selective phase
rotation, and is inherently inseparable from quantum coherence. From
the above listed marginal probabilities, one can deduce that\begin{subequations}
\begin{align}
\sum_{j}Im\left[Q_{ij}(\rho)\right] & =0\label{eq:1-1}\\
\sum_{i}Im\left[Q_{ij}(\rho)\right] & =0\label{eq:2}\\
\sum_{ij}Im\left[Q_{ij}(\rho)\right] & =0\label{eq:3}
\end{align}
 \end{subequations} 

Under this binary non-selective measurement framework, if $\rho=\rho^{\prime}$,
that is, $\rho$ is a mixed state, both correction terms of the KD
distribution are $0$, so no outliers will occur.

\subsection{\label{subsec:2-2}Values of an system observable }

With the sense of two-state vector formalism (TSVF) of quantum theory
\citep{Aharonov1964,Aharonov1988,Aharonov1990,Aharonov2008}, the
conditional expectation value of any system observable $A$ between
the pre- and post-selected states $\rho$ and $\rho_{f}$ can be defined
by

\begin{equation}
A_{T}=\frac{Tr\left(\rho_{f}A\rho\right)}{Tr\left(\rho\rho_{f}\right)}=\frac{\langle\psi_{f}\vert A\rho\vert\psi_{f}\rangle}{\langle\psi_{f}\vert\rho\vert\psi_{f}\rangle}\label{eq:20}
\end{equation}
where $\rho$ represents the density operator of preselected state
and we assume the postselection is made with the state $\vert\psi_{f}\rangle$
so that $\rho_{f}=\vert\psi_{f}\rangle\langle\psi_{f}\vert$. 

Based on the two-state vector formalism of quantum theory, considering
standard projective measurements, we have reformulated the three known
typical values of a system's observables according to the KD distribution.

\subsubsection{\label{subsec:2-2-1}Normal expectation value}

For the system state $\rho$ if we don't consider the postselection
process, then the normal expectation value of the measured system
observable $A$ under the given system state $\rho$ is given by 
\begin{align}
\langle A\rangle & =Tr\left(A\rho\right)=\sum_{i}a_{i}\rho_{ii}\label{eq:17}
\end{align}
 where $\rho_{ii}=\langle a_{i}\vert\rho\vert a_{i}\rangle=p(a_{i}\vert\rho)$
is diagonal element of system density operator on the Hilbert space
with respect to the reference basis $\{\vert a_{i}\rangle\}$. In
general, an arbitrary density matrix $\rho$ can be expanded as:

\begin{equation}
\mathcal{\mathtt{\rho}}=\sum_{i}p(a_{i}\vert\rho)\vert a_{i}\rangle\langle a_{i}\vert+\sum_{i\neq k}\langle a_{i}\vert\rho\vert a_{k}\rangle\vert a_{i}\rangle\langle a_{k}\vert\label{eq:18}
\end{equation}
 As we can see only the incoherent part $\sum_{i}p(a_{i}\vert\rho)\vert a_{i}\rangle\langle a_{i}\vert$
of the density matrix of $\rho$ contributed to measurement result
and missing the off-diagonal elements, and it belong the strong measurement.
In this measurement scenario, as mentioned, not exist postselect measurement
process. Hence, by using the Eq. (\ref{eq:5}), we can express the
expectation value $\langle A\rangle$ in terms of KD distribution
as: 
\begin{align}
\langle A\rangle & =\sum_{i}a_{i}p(a_{i}\vert\rho)=\sum_{ij}a_{i}Q_{winger}.\label{eq:19}
\end{align}

\subsubsection{\label{subsec:2-2-2}Conditional expectation value }

\begin{figure}
\includegraphics[width=8cm]{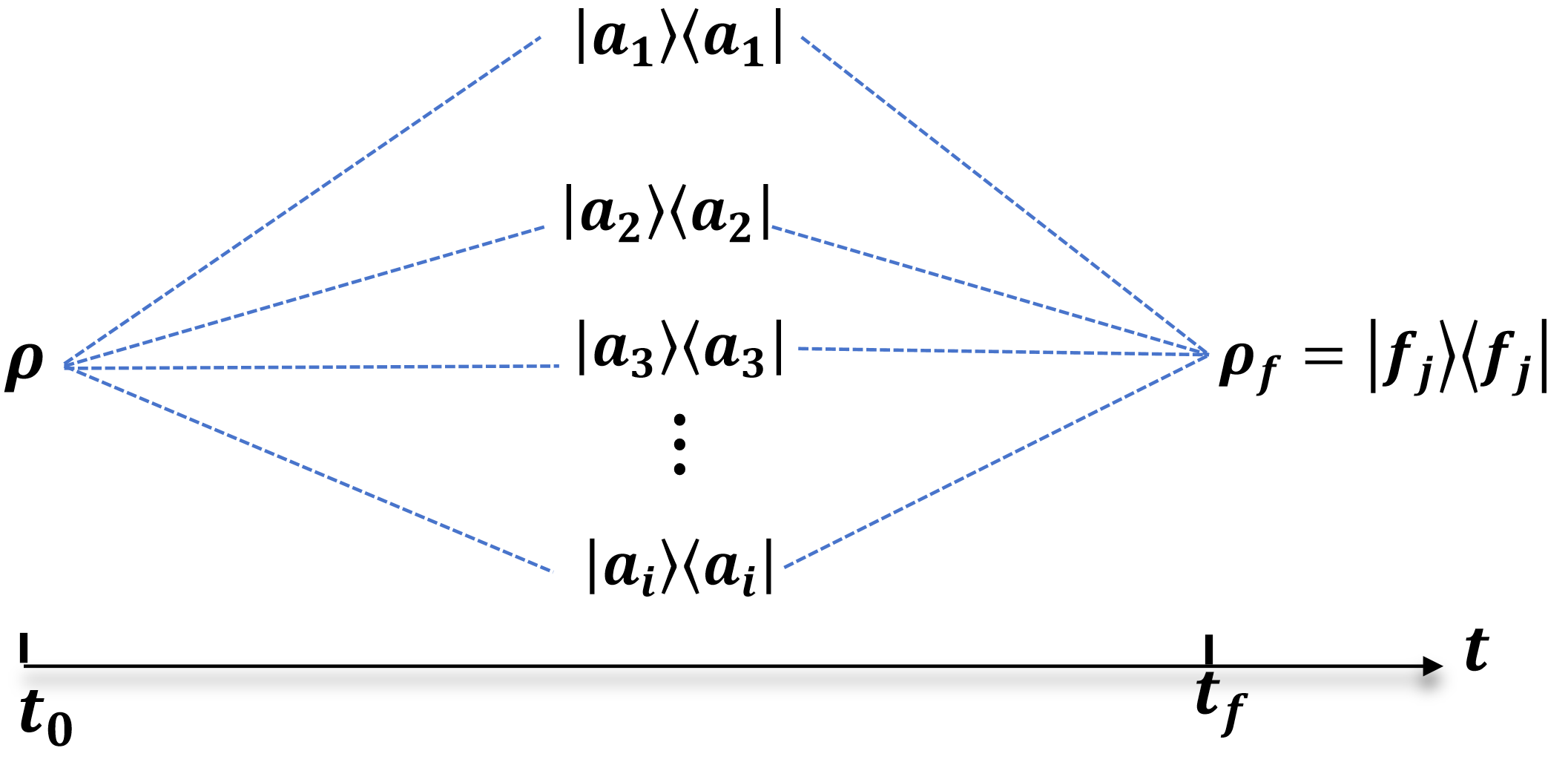}

\caption{\label{fig:1-1}Schematics of conditional expectation value and weak
value . According to the TSVF formalism of the quantum theory the
system state prepared in $\rho$ at $t_{0}$ and postselected to $\vert f_{j}\rangle$
at $t_{f}$. The expectation value of the system observable $A=\sum_{i}a_{i}\Pi_{a_{i}}$
give by $A_{c}$ or weak value $\langle A\rangle_{w}$ for different
measurement strengths.}
\end{figure}

If we consider the two-state vector formalism (TSVF) of quantum theory,
the conditional expectation value of the measured system observable
between pre- and post-selected states $\vert\psi_{i}\rangle$ and
$\vert\psi_{f}\rangle$ is given by the ABL role \citep{Aharonov1964,Ban2015,Aharonov1991}:
\begin{equation}
A_{c}=\sum_{i}a_{i}\frac{\vert\langle\psi_{f}\vert\Pi_{a_{i}}\vert\psi_{i}\rangle\vert^{2}}{\sum_{i}\vert\langle\psi_{f}\vert\Pi_{a_{i}}\vert\psi_{i}\rangle\vert^{2}}=\sum_{i}a_{i}P\left(a_{i},\vert\psi_{f}\rangle\left|\right.\vert\psi_{i}\rangle\right),\label{eq:12-1}
\end{equation}
where $P\left(a_{i},\vert\psi_{f}\rangle\left|\right.\vert\psi_{i}\rangle\right)$
is the conditional probability for finding the eigenvalue $a_{i}$
under the pre- and post-selected states $\vert\psi_{i}\rangle$ and
$\vert\psi_{f}\rangle$. The illustration of conditional measurement
characterized by TSVF is depicted in Fig. \ref{fig:1-1}. 

Next, for simplicity we let the standard projection measurement to
be $\vert\psi_{f}\rangle=\vert f_{j}\rangle$, and $\rho=\vert\psi_{i}\rangle\langle\psi_{i}\vert$
(see Fig. \ref{fig:1-1}). It can be observed that the $A_{c}$ can
be rewritten in terms of the Wigner formula $Q_{wigner}$ as: 
\begin{align}
A_{c} & =\sum_{i}a_{i}\frac{\vert\langle f_{j}\vert\Pi_{a_{i}}\vert\psi_{i}\rangle\vert^{2}}{\sum_{i}\vert\langle f_{j}\vert\Pi_{a_{i}}\vert\psi_{i}\rangle\vert^{2}}=\frac{\sum_{i}a_{i}Tr\left(\Pi_{a_{i}}\rho\Pi_{a_{i}}\Pi_{f_{j}}\right)}{\sum_{i}Tr\left(\Pi_{a_{i}}\rho\Pi_{a_{i}}\Pi_{f_{j}}\right)}\nonumber \\
 & =\frac{\sum_{i}a_{i}Q_{wigner}}{\sum_{i}Q_{wigner}}.\label{eq:21}
\end{align}
The KD distribution provides a universal mathematical framework for
the concept of conditional expectation, applicable to both mixed states
and continuous variable systems. Here we have to notice that the denominator
$\sum_{i}Q_{wigner}=Tr(\Pi_{f_{j}}\rho^{\prime})$ represents the
probability of successful postselection of the specific final state
$\vert f_{j}\rangle$ under the state $\rho^{\prime}$. If we don't
consider the postselection events, the operation of summation over
all final states in the complete basis $\{\vert f_{j}\rangle\}_{j=1}^{d}$
is taken and then the $A_{c}$ reduced to the conventional expectation
value $\langle A\rangle$ (see the Eq. (\ref{eq:19})). 

\subsubsection{\label{subsec:2-2-3}WEAK VALUE}

The concept of the weak value was introduced by Aharonov, Albert and
Vaidman \citep{Aharonov1988}. If we consider an ensemble of quantum
systems prepared in preselected state $\rho=\vert\psi_{i}\rangle\langle\psi_{i}\vert$,
and a post-selected measurement is performed which retaining only
those systems found in the state $\vert\psi_{f}\rangle$, the weak
value of the observable $A$ is defined as 

\begin{equation}
A_{w}=\frac{\langle\psi_{f}\vert A\vert\psi_{i}\rangle}{\langle\psi_{f}\vert\psi_{i}\rangle}=\frac{\langle\psi_{f}\vert A\rho\vert\psi_{f}\rangle}{\langle\psi_{f}\vert\rho\vert\psi_{f}\rangle}.
\end{equation}
 The weak value $A_{w}$ is a complex number and can lie far outside
the observable\textquoteright s eigenvalue spectrum, and have been
experimentally verified in numerous contexts \citep{Ritchie1991,Hosten2008,Aharonov2013,Xu2022,Dressel2014}.
The experimental procedure for extracting such values is known as
a weak measurement \citep{Sokolovski2013,Svensson2013}. Its real
part is interpreted as the average shift in a measurement device\textquoteright s
pointer, and its imaginary part is related to the shift in the conjugate
momentum of the pointer \citep{Jozsa2007,Dressel2014}.

By taking the postselected state $\vert\psi_{f}\rangle=\vert f_{j}\rangle$,
we can express the weak value in terms of the KD distribution, i.e:
\begin{align}
A_{w} & =\frac{\sum_{i}a_{i}\langle\psi_{f}\vert\Pi_{a_{i}}\rho\vert\psi_{f}\rangle}{\langle\psi_{f}\vert\rho\vert\psi_{f}\rangle}=\frac{\sum_{i}a_{i}Q_{ij}(\rho)}{p(f_{j}\vert\rho\text{)}}\nonumber \\
 & =\frac{\sum_{i}a_{i}Q_{ij}(\rho)}{\sum_{i}Q_{ij}(\rho)}.\label{eq:25}
\end{align}
Since the KD distribution is complex, it is a natural element to characterize
the weak value. Using KD distributions as a bridge to incorporate
weak values into a broader quantum measurement theory can explain
the behavior of weak values in the presence of decoherence, changes
in measurement strength, and mixed states. If the KD distribution
is known, the weak value of the observable can be obtained without
considering the typical weak measurement strengths. That is to say,
weak values also can be observed via projective measurements with
an indefinite causal order. In this scenario, as showed in previous
studies \citep{Arvidsson-Shukur2024}. the weak measurement is not
necessary condition to obtain the weak value within TSVF of quantum
theory.

The interesting point is that we can directly measure both the real
and imaginary parts of the weak value from the KD distribution, thereby
endowing the weak value with an operational meaning. Without employing
conventional weak measurement procedures , the weak value can be fully
reconstructed solely via standard projective measurements and phase
rotations.Next we explain the meaning of the weak value with the help
of KD distribution. The real and imaginary parts of the weak value
$\langle A\rangle_{w}$ reads as
\begin{align}
Re\langle A\rangle_{w} & =\frac{\sum_{i}a_{i}Re[Q_{ij}(\rho)]}{\sum_{i}Q_{ij}(\rho))}\nonumber \\
 & =\frac{\sum_{i}a_{i}Q_{wigner}}{\sum_{i}Re\left[Q_{ij}(\rho)\right]}+\frac{\sum_{k\neq i}a_{i}Re\left[\rho_{ki}\Psi_{ki}\right]}{\sum_{i}Re\left[Q_{ij}(\rho)\right]},\label{eq:24}
\end{align}
 and 
\begin{align}
Im\langle A\rangle_{w} & =\frac{\sum_{i}a_{i}Im\left[Q_{ij}(\rho)\right]}{\sum_{i}Q_{ij}(\rho)}\nonumber \\
 & =\frac{\sum_{k\neq i}a_{i}Im\left[\rho_{ki}\Psi_{ki}\right]}{\sum_{i}Re\left[Q_{ij}(\rho)\right]},\label{eq:25-1}
\end{align}
respectively. From this expressions we can deduce that if $\rho_{ki}\neq0$
and $\Psi_{ki}\neq0$ simultaneously, then there will occur the anomalous
values of weak value. That is to say, the anomalous weak values require
coherence under the eigenbasis of measured system observable \citep{Wagner2023}.
Anomalous weak values are a direct manifestation of quantum interference
effects. Under this condition, quantum coherence is 'hidden' in standard
measurements, but through the special mechanisms of weak measurement
and post-selection, this coherence is amplified and reflected in the
measurement results, appearing as anomalous weak values \citep{Matzkin2020}.

Among them, the Weak-value real part Wigner term $\frac{\sum_{i}a_{i}Q_{wigner}}{\sum_{i}Q_{ij}(\rho)}$
reflects the expectation under unconditional measurement. We compare
it with the conditional expectation value $A_{c}=\frac{\sum_{i}a_{i}Q_{wigner}}{\sum_{i}Q_{wigner}}$,
where$A_{c}$ includes the \textquotedblleft state update\textquotedblright ,
that is, the wave function collapse in quantum mechanics. This corresponds
to a strong measurement process, emphasizing the significant disturbance
of the measurement on the system's state and the subsequent probability
adjustment. The former corresponds to the global expectation value
under unconditional probability, reflecting the system's average behavior
over all possible measurement paths, without involving measurement-induced
quantum state collapse or local responses, This corresponds to a weak
measurement process. In other words, it describes the global average
behavior in the weak measurement process, encompassing all possible
measurement paths \citep{Duprey2017,Matzkin2020}, and therefore has
a broader statistical coverage. Although both formulas above involve
conditional probability, the essential difference lies in whether
they include the 'state update' (wave function collapse) caused by
quantum measurement. The weak-value real part Wigner term describes
the statistical behavior of the system in the absence of interference,
while $A_{c}$ takes into account the change in the system's state
caused by the measurement itself. This is the core reason for the
difference between the two. The following are the denominators of
the two respectively
\begin{equation}
\sum_{i}Q_{wigner}=Tr\left(\rho'\Pi_{f_{j}}\right)\label{eq:21-1}
\end{equation}
\begin{align}
\sum_{i}Q_{ij}(\rho) & =Tr\left(\rho\Pi_{f_{j}}\right)\nonumber \\
 & =Tr\left(\rho'\Pi_{f_{j}}\right)+Re\left[\sum_{k\neq i}\rho_{ki}\Psi_{ki}\right]\\
 & =p(f_{j}\vert\rho)\label{eq:28}
\end{align}
The conditional probability $Tr\left(\rho'\Pi_{f_{j}}\right)$ of
a strong measurement depends on the specific measurement path (i.e.,
the pre-selection outcome $a_{i}$), whereas the unconditional probability
$p(f_{j})$ of a weak measurement is an average over all possible
paths. This path dependence is a manifestation of quantum contextuality,
meaning that the measurement result depends not only on the measured
system itself but also on the entire measurement context (including
previous measurement operations).

In above we introduced three different values of measured system observable
correspond to different measurements and the connections between them
are explained in terms of the meaning of the KD distribution. In next
section we further show connections of between different measurements
types via dynamical KD distribution. 

\section{\label{sec:3}measurement transition with dynamical KD distribution }

\begin{figure}
\includegraphics[width=8cm]{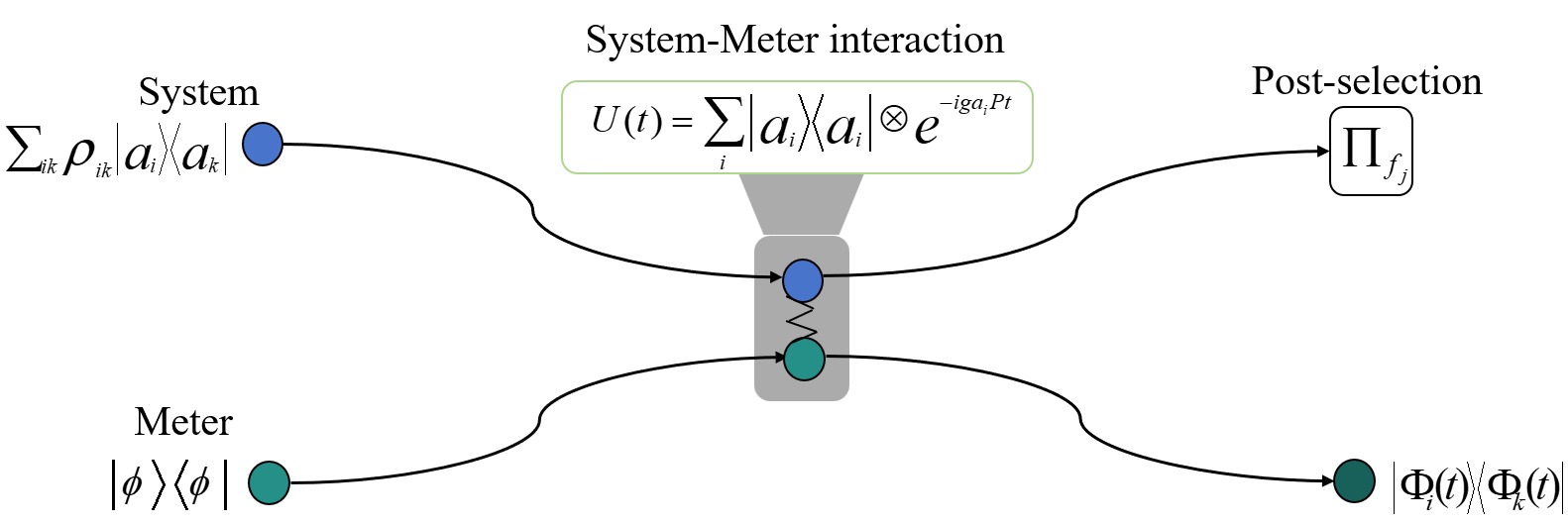}

\caption{\label{fig:1}The initial state $\rho=\sum_{ik}\rho_{ik}\vert a_{i}\rangle\langle a_{k}\vert$
of the system interacts with the initial state�$\vert\phi\rangle\langle\phi\vert$
of the pointer, and the interaction operator is $U(t)=\sum_{i}\vert a_{i}\rangle\langle a_{i}\vert\otimes U_{i}(t)$.
$A$ and $P$ are observables of the system and pointer, respectively.
After performing an operation on the system, the pointer state becomes
$\vert\phi_{i}(t)\rangle\langle\phi_{k}(t)\vert=\vert\phi\left(x-ga_{i}t\right)\rangle\langle\phi\left(x-ga_{k}t\right)\vert$.}

\end{figure}

A standard measurement issue consisted three parts: measured system,
pointer ( also called measuring device/measurement apparatus ) and
environment which realize the interaction between measured system
and pointer\citep{Jacobs2014}. Hence, the form of the total Hamiltonian
of a measurement problem can be written as $H=H_{0}+H_{int}$. Herein,
the $H_{0}$ is the free Hamiltonian of the pointer plus measured
system and $H_{int}$ represents the interaction part. For standard
measurement problems the interaction between them is characterized
by von Neumann type Hamiltonian $H_{int}=gA\otimes Q$. Here, $A$
is measured system observable, $g$ denotes the interaction strength
and the pointer observable $Q$ is defined as a canonical quadrature,
either position $X$ or momentum $P$ operator. For ideal measurement
it requires $[H,A]=\left[H_{0},A\right]=0$ and mean energy of the
measured system also constant in time. For simplicity we assume that
the total Hamiltonian is
\begin{equation}
H=\frac{p^{2}}{2m}+gA\otimes P.\label{eq:34-1}
\end{equation}
For the initially prepared system state $\rho=\vert\psi_{i}\rangle\langle\psi_{i}\vert=\sum_{ik}\rho_{ik}\vert a_{i}\rangle\langle a_{k}\vert$
and Gaussian profile 
\begin{equation}
\vert\phi\rangle=\left(\frac{1}{2\pi\sigma^{2}}\right)^{\frac{1}{4}}\int exp\left(-\frac{x^{2}}{4\sigma^{2}}\right)dx\vert x\rangle\label{eq:35}
\end{equation}
of the pointer the evolution operator can be decomposed into sub-evolution
operators corresponding to the system's eigenstates:

\begin{equation}
U(t)=\sum_{i}\vert a_{i}\rangle\langle a_{i}\vert\otimes U_{i}(t).\label{eq:36}
\end{equation}
This schematics of the this measurement process is shown in Fig. \ref{fig:1}.
Under this unitary time evolution operator, the composed initial state
$\Phi(0)=\rho\otimes\vert\phi\rangle\langle\phi\vert$ of total system
changed to $\Phi(t)$
\begin{align}
\Phi(t) & =U(t)\Phi(0)U^{\dagger}(t)\nonumber \\
 & =\sum_{ik}\rho_{ik}\vert a_{i}\rangle\langle a_{k}\vert\otimes\vert\phi_{i}(t)\rangle\langle\phi_{k}(t)\vert.\label{eq:67}
\end{align}
Here, 
\begin{equation}
\vert\phi_{i}(t)\rangle=U_{i}(t)\vert\phi\rangle=e^{-iga_{i}Pt}\vert\phi\rangle=\vert\phi(x-gta_{i})\rangle\label{eq:70}
\end{equation}
 is the final state of the pointer. This indicated that the center
of the pointer wave packet shifted from the initial $x_{0}=0$ point
to $x_{0i}=ga_{i}t$, with the displacement proportional to the system's
eigenvalue $a_{i}$ and the interaction parameter $gt$, effectively
'copying' the system state onto the pointer's position. It can see
that after the time evolution the measured system and pointer entangled
and the final pointer carries the system information we want to read.
As a result the initially prepared pure state $\rho$ of the measured
system become changed and its spoiled degree depends on the interaction
strength $g$ and length of interaction time $t$. 

Our main concern is the disturbance of measurement apparatus to system
state during the measurement process. Furthermore, the reduced density
matrix $\rho(t)$ of measured system after the measurement is obtained
by tracing out the degrees of freedom of the pointer
\begin{align}
\rho(t) & =Tr_{P}\left[\Phi(t)\right]\nonumber \\
 & =\sum_{i}\rho_{ii}\vert a_{i}\rangle\langle a_{i}\vert+\sum_{i\neq k}\rho_{ik}\vert a_{i}\rangle\langle a_{k}\vert\langle\phi_{i}(t)\vert\phi_{k}(t)\rangle.
\end{align}
It represents the system state after measurement associated with the
eigenvalue $a_{k}$ of the $A$ which corresponding the the eigenstate
$\vert a_{k}\rangle$ and the overlap integral of the pointer state
is given by 
\begin{align}
\langle\phi_{i}(t)\vert\phi_{k}(t)\rangle & =\int\phi^{*}\left(x-ga_{i}t\right)\phi\left(x-ga_{k}t\right)dx\nonumber \\
 & =exp\left[-\frac{\left(\bigtriangleup x\right)^{2}}{8\sigma^{2}}\right]\label{eq:over}
\end{align}
Here, $\bigtriangleup x=g\left(a_{i}-a_{k}\right)t$ represents the
distance between the centers of the two wave packets. In this case,
it is clear that the overlap integral of the pointer states is real
and it can characterized existence of the quantum interference of
$\rho(t)$ in the basis $\{\vert a_{i}\rangle\}_{j=1}^{d}$. The overlap
$\langle\phi_{i}(t)\vert\phi_{k}(t)\rangle$ of between $\left|\phi_{i}(t)\right\rangle $
and $\left|\phi_{k}(t)\right\rangle $ serves as a quantitative measure
delineating the degree of interference \citep{Schlosshauer2019},
and it refers the decoherence function \citep{Breuer2007}
\begin{equation}
F(t)=\left|\left\langle \phi_{i}(t)\mid\phi_{k}(t)\right\rangle \right|.\ \label{eq:11-1}
\end{equation}
The decoherence function $F(t)$ is a function that changes over time
and describes the process by which coherence (non-diagonal) gradually
decays over time. It is a non-negative time-dependent function, known
as the or \textquotedbl decay factor\textquotedbl . The strength
of quantum coherence decays with the time exponent. In the theory
of decoherence, the overlapping integral describes the \textquotedbl coincidence\textquotedbl{}
of two Gaussian wave packets, and its mode length directly reflects
the strength of quantum coherence \citep{Joos1985,Breuer2007,Schlosshauer2005,Brunner2003}. 

This function can describe the behavior of the nondiagonal elements
of the reduced density matrix $\rho(t)$ when $i\ne k$. In order
to describe the decreasing the $F(t)$ quantitatively, we introduce
a term called \textit{decoherence time} $\tau_{D}=8\sigma^{2}/g^{2}(\triangle x)^{2}$.
For $i\neq k$ the overlap of the states $\vert\phi_{i}(t)\rangle$
and $\left|\phi_{k}(t)\right\rangle $ approaches to zero after times
which are large compared to a typical time scale $\tau_{D}$ of the
system such that 
\begin{equation}
\langle\phi_{j}(t)\vert\phi_{i}(t)\rangle\rightarrow\delta_{ij},\ \ for\ \ t\gg\tau_{D},
\end{equation}
then the reduced density matrix of the system became as
\begin{equation}
\rho(t)\to\sum\limits_{i}\rho_{ii}\left|a_{i}\right\rangle \left\langle a_{i}\right|.\label{eq:12-1-1}
\end{equation}
 This result showed that the coherences of density matrix of our system
has vanished after long time ($t\gg\tau_{D}$) interaction with the
pointer. In the measurement problem, after $t\gg\tau_{D}$ the state
$\rho(t)$ of the system behaves as a incoherent mixture of the state
$\vert a_{i}\rangle$ so that in the conventional expectation value
the interference terms of the form $\langle a_{i}\vert A\vert a_{j}\rangle$
($i\neq j$ ) of any system observable $A$ are no longer occur. 

The dynamical time evolution which characterized by decoherence function
of the system state can help us to deeply understand the measurement
transition in terms time dependent KD distribution. By using the Eq.
(\ref{eq:20}), we can obtain the general expectation value of measured
system observable $A$ for pre- and post-selected states $\rho(t)$
and $\vert f_{j}\rangle$ as 
\begin{equation}
A_{T}=\frac{\langle f_{j}\vert A\rho(t)\vert f_{j}\rangle}{\langle f_{j}\vert\rho(t)\vert f_{j}\rangle}=\frac{\sum_{i}a_{i}Q_{ij}(\rho(t),F)}{\sum_{i}Q_{ij}(\rho(t),F)},\label{eq:33}
\end{equation}
where the time dependent KD distribution reads as
\begin{align}
Q_{ij}(\rho(t),F) & =Tr\left(\Pi_{a_{i}}\rho\left(t\right)\Pi_{f_{j}}\right)\nonumber \\
 & =FQ_{ij}(\rho)+\left(1-F\right)Q_{wigner}\label{eq:90}
\end{align}
For the details of this derivation, please see Appendix A. The quantity
$A_{T}$ can represent the general expectation value of the system
observable which is valid in entire measurement regimes described
by TSVF in quantum theory. The decoherence fucntion $F$ bound to
$\left[0,1\right]$. For one extreme case, i.e., $F=0$, which belong
to strong conditional strong measurement, the time dependent KD distribution
$Q_{ij}(\rho(t),F)$ reduced to the Winger formula $Q_{winger}$.
Consequently, the $A_{T}$ gives us the conditional expectation value
$A_{c}$: 
\begin{equation}
\left(A_{T}\right)_{F\rightarrow0}=\frac{\sum_{i}a_{i}Q_{wigner}}{\sum_{i}Q_{wigner}}=A_{c}.\label{eq:56}
\end{equation}
 On the other extreme case, $F=1$, the time dependent KD distribution
gives the normal time independent KD distribution $Q_{ij}(\rho)=Q(a_{i},f_{j}\vert\rho)$
and $A_{T}$ reduced to the weak value, i.e., 
\begin{equation}
\left(A_{T}\right)_{F\rightarrow1}=\frac{\sum_{i}a_{i}Q_{ij}(\rho)}{\sum_{i}Q_{ij}(\rho)}=A{}_{w}.\label{eq:43}
\end{equation}
 Furthermore, the intermediate regions, $F\in(0,1)$, are important
to describe the continuous measurement transition processes in terms
of KD distribution. Thus, our dynamical KD framework not only fills
the gap between weak and strong measurements, but also provides a
clear, physically motivated pathway---through pointer-induced decoherence---for
understanding how quasi-probabilities evolve with measurement strength.
In this way, the KD distribution is shown to be the natural and complete
mathematical structure for describing conditioned measurement sequences
at all interaction strengths, smoothly interpolating from anomalous
weak values to classical projective probabilities.

\section{\label{sec:4}Non-classical quantization of KD distribution}

The non-classicality of the KD distribution, manifested through its
negativity and non-reality, has been quantified by various measures
to distinguish quantum behavior from classical statistics \citep{DeBievre2021,Xu2022,Langrenez2023,DeBievre2023}.
A prominent measure, introduced to capture the total magnitude of
quantum corrections, is given by \citep{He2024} 
\begin{align}
\mathcal{N}[\rho(0)] & =\frac{1}{2}\sum_{i,j}\left|Tr\left((\rho-\rho')\Pi_{f_{j}}\right)\right|\nonumber \\
 & +\frac{1}{2}\sum_{i,j}\left|Tr\left((\rho-\rho')\Pi_{f_{j}}^{\frac{\pi}{2}}\right)\right|.\label{eq:42}
\end{align}
While previous studies have established $\mathcal{N}[\rho]$ as a
faithful witness of KD non-classicality for static states, our framework
reveals its dynamical evolution under measurement-induced decoherence---providing
a quantitative link between the loss of quantumness and the continuous
tuning of measurement strength. 

The key to understanding this evolution lies in combining the structure
of $\mathcal{N}[\rho]$ with our central result from Sec. \ref{sec:3}---the
time-dependent KD distribution $Q_{ij}(\rho(t),F)$ which can be rewitten
as 
\begin{equation}
Q_{ij}(\rho(t),F)=Q_{wigner}+F\triangle Q_{ij},\label{eq:43-1}
\end{equation}
 Herein, $\Delta Q_{ij}=Q_{ij}(\rho)-Q_{\text{wigner}}$ represent
the non-classical part of the initial KD distribution. This reveals
that the classical component remains invariant under decoherence,
while quantum corrections are uniformly attenuated by the factor $F$.Decoherence
only diminishes the non-classical characteristics of a quantum system,
while it has no effect on classical statistical behavior $Q_{wigner}$
initially.

Since the measure $\mathcal{N}[\rho(0)]$ depends linearly on the
magnitude of $\triangle Q_{ij}$ (as evident from its definition),
we obtain the fundamental dynamical relation: 

\begin{equation}
\mathcal{N}[\rho(t)]=F\mathcal{N}[\rho(0)].\label{eq:44}
\end{equation}
 Equation (\ref{eq:44}) establishes a quantitative bridge between
the formal measurement-strength parameter $F$ and the physical quantumness
resource $\mathcal{N}$ , with several important implications:

1. Direct answer to the central question: This relation directly addresses
when the KD description is most appropriate: it remains essential
precisely when $\mathcal{N}$ is significant, i.e., in regimes where
quantum coherence actively influences measurement statistics. The
continuous decay of $\mathcal{N}$ with $F$ maps directly onto the
smooth transition from weak to strong measurement regimes.

2. Predictive power for non-classical phenomena: The decay of $\mathcal{N}$
quantitatively governs the feasibility of observing phenomena like
anomalous weak values. A significant $\mathcal{N}$ is a necessary
condition for weak values to lie outside an observable's eigenvalue
spectrum. As $F$ decreases, the shrinking $\mathcal{N}$ predicts
the attenuation of such anomalies, providing a theoretical tool to
estimate the measurement strength required to suppress specific quantum
effects.

2. Experimental testability: For typical decoherence channels where
$F=e^{-t/\tau_{D}}$ ( with $\tau_{D}$ the decoherence time), Eq.
(\ref{eq:44}) predicts $\mathcal{N}[\rho(t)]=e^{-t/\tau_{D}}\mathcal{N}[\rho(0)]$.
This offers an experimentally testable signature: the non-classicality
measure which can be reconstructed from standard projective measurements
and phase rotations as discussed in Sec. \ref{sec:2}---should decay
at the same rate as the pointer overlap $F(t)$.

The dynamical relation (\ref{eq:44}) completes our unified framework
by integrating the kinematic and dynamic perspectives on the measurement
transition:

Kinematic ($F$) : Governs the formal transformation of the KD distribution
from a complex quasiprobability to a real, non-negative one.

Dynamic ($\mathcal{N}$): Q uantifies the physical quantumness that
is progressively lost during this transformation.

This dual description establishes the KD distribution not merely as
a mathematical connector between measurement paradigms, but as a central
physical object whose evolving non-classical content directly mirrors
the process of decoherence. The linear decay law confirms that pointer-induced
decoherence is the universal mechanism governing both the formal interpolation
between regimes and the quantitative suppression of quantum features
across the entire measurement landscape.

\section{\label{sec:5}conclusion }

We have established the KD quasiprobability as the universal mathematical
framework that seamlessly connects all quantum measurement regimes.
The framework naturally unifies normal expectation values, conditional
expectation values, and weak values through a single informationally
complete representation.

Our analysis reveals pointer-induced decoherence as the fundamental
physical mechanism governing the continuous transition between measurement
strengths. The decoherence factor quantitatively interpolates between
weak and strong measurements, explaining how quasiprobabilities transform
from complex-valued distributions to classical joint probabilities.
This provides a direct answer to when the KD description is most appropriate:
precisely in regimes where quantum coherence significantly influences
measurement statistics.

These results resolve the longstanding question of the KD distribution's
applicability by establishing it as the central structure from which
the phenomenology of all quantum measurements emerges through a physically
transparent decoherence pathway.
\begin{acknowledgments}
This work was supported by the National Natural Science Foundation
of China (No. 12365005).
\end{acknowledgments}

\bibliographystyle{apsrev4-1}
\bibliography{Reference}

@article{Johansen2007,
  title = {Quantum theory of successive projective measurements},
  author = {Johansen, Lars M.},
  journal = {Phys. Rev. A},
  volume = {76},
  issue = {1},
  pages = {012119},
  numpages = {6},
  year = {2007},
  month = {Jul},
  publisher = {American Physical Society},
  doi = {10.1103/PhysRevA.76.012119},
  url = {https://link.aps.org/doi/10.1103/PhysRevA.76.012119}
}

@article{He2024,
  title = {Nonclassicality of the Kirkwood-Dirac quasiprobability distribution via quantum modification terms},
  author = {He, Jiayu and Fu, Shuangshuang},
  journal = {Phys. Rev. A},
  volume = {109},
  issue = {1},
  pages = {012215},
  numpages = {12},
  year = {2024},
  month = {Jan},
  publisher = {American Physical Society},
  doi = {10.1103/PhysRevA.109.012215},
  url = {https://link.aps.org/doi/10.1103/PhysRevA.109.012215}
}

@article{Song2024,
  title = {Theoretical investigation of the relations between pointer-induced quantum decoherence and the weak-to-strong measurement transition},
  author = {Song, Xiao-Feng and Liu, Shuang and Chen, Xi-Hao and Turek, Yusuf},
  journal = {Phys. Rev. A},
  volume = {110},
  issue = {6},
  pages = {062217},
  numpages = {19},
  year = {2024},
  month = {Dec},
  publisher = {American Physical Society},
  doi = {10.1103/PhysRevA.110.062217},
  url = {https://link.aps.org/doi/10.1103/PhysRevA.110.062217}
}

@article{Aharonov1988,
  title = {How the result of a measurement of a component of the spin of a spin-1/2 particle can turn out to be 100},
  author = {Aharonov, Yakir and Albert, David Z. and Vaidman, Lev},
  journal = {Phys. Rev. Lett.},
  volume = {60},
  issue = {14},
  pages = {1351--1354},
  numpages = {0},
  year = {1988},
  month = {Apr},
  publisher = {American Physical Society},
  doi = {10.1103/PhysRevLett.60.1351},
  url = {https://link.aps.org/doi/10.1103/PhysRevLett.60.1351}
}

@book{Neumann1932,
  title={Mathematical Foundations of Quantum Mechanics},
  author={J. V. Neumann},
  year={1932},
  publisher = {Springer Press, Berlin},
}

@article{Kirkwood1933,
  title = {Quantum Statistics of Almost Classical Assemblies},
  author = {Kirkwood, John G.},
  journal = {Phys. Rev.},
  volume = {44},
  issue = {1},
  pages = {31--37},
  numpages = {0},
  year = {1933},
  month = {Jul},
  publisher = {American Physical Society},
  doi = {10.1103/PhysRev.44.31},
  url = {https://link.aps.org/doi/10.1103/PhysRev.44.31}
}

@article{Dirac1945,
  title = {On the Analogy Between Classical and Quantum Mechanics},
  author = {Dirac, P. A. M.},
  journal = {Rev. Mod. Phys.},
  volume = {17},
  issue = {2-3},
  pages = {195--199},
  numpages = {0},
  year = {1945},
  month = {Apr},
  publisher = {American Physical Society},
  doi = {10.1103/RevModPhys.17.195},
  url = {https://link.aps.org/doi/10.1103/RevModPhys.17.195}
}

@article{Pusey2014,
  title = {Anomalous Weak Values Are Proofs of Contextuality},
  author = {Pusey, Matthew F.},
  journal = {Phys. Rev. Lett.},
  volume = {113},
  issue = {20},
  pages = {200401},
  numpages = {5},
  year = {2014},
  month = {Nov},
  publisher = {American Physical Society},
  doi = {10.1103/PhysRevLett.113.200401},
  url = {https://link.aps.org/doi/10.1103/PhysRevLett.113.200401}
}

@article{Dressel2014,
  title = {Colloquium: Understanding quantum weak values: Basics and applications},
  author = {Dressel, Justin and Malik, Mehul and Miatto, Filippo M. and Jordan, Andrew N. and Boyd, Robert W.},
  journal = {Rev. Mod. Phys.},
  volume = {86},
  issue = {1},
  pages = {307--316},
  numpages = {10},
  year = {2014},
  month = {Mar},
  publisher = {American Physical Society},
  doi = {10.1103/RevModPhys.86.307},
  url = {https://link.aps.org/doi/10.1103/RevModPhys.86.307}
}

@article{Wagner2023,
  title = {Simple proof that anomalous weak values require coherence},
  author = {Wagner, Rafael and Galv\~ao, Ernesto F.},
  journal = {Phys. Rev. A},
  volume = {108},
  issue = {4},
  pages = {L040202},
  numpages = {6},
  year = {2023},
  month = {Oct},
  publisher = {American Physical Society},
  doi = {10.1103/PhysRevA.108.L040202},
  url = {https://link.aps.org/doi/10.1103/PhysRevA.108.L040202}
}

@article{Margenau1961,
    author = {Margenau, Henry and Hill, Robert Nyden},
    title = {Correlation between Measurements in Quantum Theory: },
    journal = {Progress of Theoretical Physics},
    volume = {26},
    number = {5},
    pages = {722-738},
    year = {1961},
    month = {11},
    issn = {0033-068X},
    doi = {10.1143/PTP.26.722},
    url = {https://doi.org/10.1143/PTP.26.722},
   }

@article{Arvidsson-Shukur2024,
doi = {10.1088/1367-2630/ada05d},
url = {https://doi.org/10.1088/1367-2630/ada05d},
year = {2024},
month = {dec},
publisher = {IOP Publishing},
volume = {26},
number = {12},
pages = {121201},
author = {Arvidsson-Shukur, David R M and Braasch Jr, William F and De Bièvre, Stephan and Dressel, Justin and Jordan, Andrew N and Langrenez, Christopher and Lostaglio, Matteo and Lundeen, Jeff S and Halpern, Nicole Yunger},
title = {Properties and applications of the Kirkwood–Dirac distribution},
journal = {New Journal of Physics}
}

@article{Hofmann2012,
doi = {10.1088/1367-2630/14/4/043031},
url = {https://doi.org/10.1088/1367-2630/14/4/043031},
year = {2012},
month = {apr},
publisher = {IOP Publishing},
volume = {14},
number = {4},
pages = {043031},
author = {Hofmann, Holger F},
title = {Complex joint probabilities as expressions of reversible transformations in quantum mechanics},
journal = {New Journal of Physics}
}

@article{Johansen2004,
    title = {Weak measurements with arbitrary probe states (Weak Measurements with Arbitrary Pointer States)},
    author = {Johansen, Lars M.},
    journal = {Physical Review Letters},
    volume = {93},
    number = {12},
    pages = {120402},
    year = {2004},
    month = {Sep},
    publisher = {American Physical Society},
    doi = {10.1103/PhysRevLett.93.120402},
    url = {https://link.aps.org/doi/10.1103/PhysRevLett.93.120402},
    archivePrefix = {arXiv},
    eprint = {quant-ph/0402050}
}

@article{Lueders1951,
    author = {Lüders, Gerhard},
    title = {Über die Zustandsänderung durch den Messprozeß},
    journal = {Ann. Phys. (Leipzig)},
    year = {1951},
    volume = {8},
    number = {1-2},
    pages = {322--328},
    doi = {10.1002/andp.19515350305},
   }

@article{Johansen2008,
    author = {Johansen, Lars M. and Mello, Pier A.},
    title = {Quantum Mechanics of Successive Measurements with Arbitrary Meter Coupling},
    journal = {Phys. Lett. A},
    volume = {372},
    pages = {5760--5764},
    year = {2008},
    doi = {10.1016/j.physleta.2008.07.039}
}

@article{Rastegin2025,
    author = {Rastegin, Alexey E.},
    title = {Kirkwood–Dirac Quasiprobabilities for Measurements Related to Mutually Unbiased Equiangular Tight Frames},
    journal = {APL Quantum},
    volume = {2},
    pages = {026108},
    year = {2025},
    doi = {10.1063/5.0254873}
}

@article{Scott2006,
    author = {Scott, A. J.},
    title = {Mutually Unbiased Bases},
    journal = {Contemp. Phys.},
    volume = {48},
    pages = {95--110},
    year = {2006},
    doi = {10.1080/00107510600960990}
}

@article{Wigner1963,
    author = {Wigner, Eugene P.},
    title = {The Problem of Measurement},
    journal = {Am. J. Phys.},
    volume = {31},
    pages = {6--15},
    year = {1963},
    doi = {10.1119/1.1969921},
    publisher = {American Association of Physics Teachers}
}

@article{Terletsky1937,
    author = {Terletsky, Y. P.},
    title = {The limiting transition from quantum to classical mechanics},
    journal = {J. Exp. Theor. Phys.},
    volume = {7},
    pages = {1290},
    year = {1937},
    publisher = {Pleiades Publishing}
}

@article{DeBievre2021,
    author = {De Bièvre, S.},
    title = {Complete incompatibility, support uncertainty, and Kirkwood-Dirac nonclassicality},
    journal = {Phys. Rev. Lett.},
    volume = {127},
    pages = {190404},
    year = {2021},
    doi = {10.1103/PhysRevLett.127.190404},
    publisher = {American Physical Society}
}

@article{Streltsov2017,
    author = {Streltsov, Alexander and Adesso, Gerardo and Plenio, Martin B.},
    title = {Colloquium: Quantum coherence as a resource},
    journal = {Rev. Mod. Phys.},
    volume = {89},
    pages = {041003},
    year = {2017},
    doi = {10.1103/RevModPhys.89.041003},
    publisher = {American Physical Society}
}

@article{Joos1985,
    author = {Joos, E. and Zeh, H. D.},
    title = {The emergence of classical properties through interaction with the environment},
    journal = {Z. Phys. B},
    volume = {59},
    pages = {223--243},
    year = {1985},
    doi = {10.1007/BF01725541},
}

@article{Schlosshauer2005,
    author = {Schlosshauer, M.},
    title = {Decoherence, the measurement problem, and the interpretation of quantum mechanics},
    journal = {Rev. Mod. Phys.},
    volume = {76},
    pages = {1267--1305},
    year = {2005},
    doi = {10.1103/RevModPhys.76.1267},
}

@article{Brunner2003,
    author = {Brunner, N. and Simon, C. and Gisin, N. and Weihs, G.},
    title = {Experimental demonstration of decoherence in a quantum interference experiment},
    journal = {Phys. Rev. Lett.},
    volume = {91},
    pages = {210402},
    year = {2003},
    doi = {10.1103/PhysRevLett.91.210402},
}

@book{Schlosshauer2007,
    author = {Schlosshauer, M.},
    title = {Decoherence and the Quantum-To-Classical Transition},
    publisher = {Springer},
    address = {Berlin},
    year = {2007},
    doi = {10.1007/978-3-540-35773-4}
}

@article{Schlosshauer2019,
    author = {Schlosshauer, M.},
    title = {Quantum decoherence},
    journal = {Phys. Rep.},
    volume = {831},
    pages = {1--57},
    year = {2019},
    doi = {10.1016/j.physrep.2019.06.001},
    publisher = {Elsevier},
}

@book{Breuer2007,
    author = {Breuer, H.-P. and Petruccione, F.},
    title = {The Theory of Open Quantum Systems},
    publisher = {Oxford University Press},
    address = {Oxford},
    year = {2007},
    doi = {10.1093/acprof:oso/9780198520634.001.0001},
}

@article{Aharonov1964,
    author = {Aharonov, Y. and Bergmann, P. G. and Lebowitz, J. L.},
    title = {Time Symmetry in the Quantum Process of Measurement},
    journal = {Phys. Rev.},
    volume = {134},
    pages = {B1410--B1416},
    year = {1964},
    doi = {10.1103/PhysRev.134.B1410},
    publisher = {American Physical Society},
}

@article{Aharonov1990,
    author = {Aharonov, Y. and Vaidman, L.},
    title = {Properties of a Quantum System During the Time Interval Between Two Measurements},
    journal = {Phys. Rev. A},
    volume = {41},
    pages = {11--20},
    year = {1990},
    doi = {10.1103/PhysRevA.41.11},
    publisher = {American Physical Society},
}

@incollection{Aharonov2008,
    author = {Aharonov, Y. and Vaidman, L.},
    title = {Weak Measurements},
    booktitle = {Time in Quantum Mechanics},
    editor = {Bassi, A. and Ghirardi, G. C.},
    publisher = {Springer},
    address = {Berlin},
    pages = {399--447},
    year = {2008},
    doi = {10.1007/978-3-540-85303-6_11}
}

@article{Ban2015,
    author = {Ban, M.},
    title = {Conditional average in a quantum system with postselection},
    journal = {Quantum Stud.: Math. Found.},
    volume = {2},
    pages = {263--273},
    year = {2015},
    doi = {10.1007/s40509-015-0034-5},
    publisher = {Springer}
}

@article{Aharonov1991,
    author = {Aharonov, Y. and Vaidman, L.},
    title = {Complete description of a quantum system at a given time},
    journal = {J. Phys. A: Math. Gen.},
    volume = {24},
    pages = {2315--2328},
    year = {1991},
    doi = {10.1088/0305-4470/24/10/024},
    publisher = {IOP Publishing}
}

@article{Jozsa2007,
    author = {Jozsa, R.},
    title = {Complex weak values in quantum measurement},
    journal = {Phys. Rev. A},
    volume = {76},
    pages = {044103},
    year = {2007},
    doi = {10.1103/PhysRevA.76.044103},
    publisher = {American Physical Society}
}

@article{Sokolovski2013,
    author = {Sokolovski, D.},
    title = {Are the Weak Measurements Really Measurements?},
    journal = {Quanta},
    volume = {2},
    pages = {50--61},
    year = {2013},
    doi = {10.12743/quanta.v2i1.15},
    publisher = {Quanta Publishing}
}

@article{Svensson2013,
    author = {Svensson, B. E. Y.},
    title = {Pedagogical Review of Quantum Measurement Theory with an Emphasis on Weak Measurements},
    journal = {Quanta},
    volume = {2},
    pages = {18--49},
    year = {2013},
    doi = {10.12743/quanta.v2i1.12},
    publisher = {Quanta Publishing}
}

@article{Ritchie1991,
    author = {Ritchie, N. W. M. and Story, J. G. and Hulet, R. G.},
    title = {Realization of a Measurement of a "Weak Value"},
    journal = {Phys. Rev. Lett.},
    volume = {66},
    pages = {1107--1110},
    year = {1991},
    doi = {10.1103/PhysRevLett.66.1107},
    publisher = {American Physical Society},
    url = {http://atomcool.rice.edu/static/ritchie_prl_1991.pdf}
}

@article{Hosten2008,
    author = {Hosten, O. and Kwiat, P.},
    title = {Observation of the Spin Hall Effect of Light via Weak Measurements},
    journal = {Science},
    volume = {319},
    number = {5864},
    pages = {787--790},
    year = {2008},
    doi = {10.1126/science.1152697},
    publisher = {American Association for the Advancement of Science}
}

@article{Xu2022,
    author = {Xu, K. and Hu, X.-M. and Zhang, C. and Huang, Y.-F. and Liu, B.-H. and Li, C.-F. and Guo, G.-C. and Hu, M.-J. and Zhang, Y.-S.},
    title = {Measuring Small Longitudinal Phase Shifts via Weak Measurement Amplification},
    journal = {Phys. Rev. A},
    volume = {105},
    pages = {042611},
    year = {2022},
    doi = {10.1103/PhysRevA.105.042611},
    publisher = {American Physical Society}
}

@article{Aharonov2013,
    author = {Aharonov, Y. and Popescu, S. and Rohrlich, D. and Skrzypczyk, P.},
    title = {Quantum Cheshire Cats},
    journal = {Phys. Rev. Lett.},
    volume = {110},
    pages = {110402},
    year = {2013},
    doi = {10.1103/PhysRevLett.110.110402},
    publisher = {American Physical Society}
}

@book{Jacobs2014,
    author = {Jacobs, K.},
    title = {Quantum Measurement Theory and its Applications},
    publisher = {Cambridge University Press},
    address = {Cambridge},
    year = {2014},
    isbn = {9781139179027},
    doi = {10.1017/CBO9781139179027},
}

@article{Langrenez2023,
    author = {Langrenez, C. and Arvidsson-Shukur, D. R. M. and De Bièvre, S.},
    title = {Characterizing the Geometry of the Kirkwood-Dirac Positive States},
    journal = {arXiv e-prints},
    eprint = {2306.00086},
    archivePrefix = {arXiv},
    primaryClass = {quant-ph},
    year = {2023},
    note = {Submitted: 31 May 2023}
}

@article{DeBievre2023,
    author = {De Bièvre, S.},
    title = {Relating Incompatibility, Noncommutativity, Uncertainty, and Kirkwood–Dirac Nonclassicality},
    journal = {J. Math. Phys.},
    volume = {64},
    pages = {022202},
    year = {2023},
    doi = {10.1063/5.0110267},
    publisher = {AIP Publishing}
}

@article{Arvidsson-Shukur2021,
    author = {Arvidsson-Shukur, D. R. M. and Drori, J. C. and Yunger Halpern, N.},
    title = {Conditions Tighter Than Noncommutation Needed for Nonclassicality},
    journal = {J. Phys. A: Math. Theor.},
    volume = {54},
    pages = {284001},
    year = {2021},
    doi = {10.1088/1751-8121/ac0289},
    publisher = {IOP Publishing}
}

@article{Duprey2017,
  title = {Null weak values and the past of a quantum particle},
  author = {Duprey, Q. and Matzkin, A.},
  journal = {Phys. Rev. A},
  volume = {95},
  issue = {3},
  pages = {032110},
  numpages = {9},
  year = {2017},
  month = {Mar},
  publisher = {American Physical Society},
  doi = {10.1103/PhysRevA.95.032110},
  url = {https://link.aps.org/doi/10.1103/PhysRevA.95.032110}
}

@article{Matzkin2020,
  title = {Weak values from path integrals},
  author = {Matzkin, A.},
  journal = {Phys. Rev. Res.},
  volume = {2},
  issue = {3},
  pages = {032048},
  numpages = {5},
  year = {2020},
  month = {Aug},
  publisher = {American Physical Society},
  doi = {10.1103/PhysRevResearch.2.032048},
  url = {https://link.aps.org/doi/10.1103/PhysRevResearch.2.032048}
}

\end{document}